\documentclass[english,pre,twocolumn,floatfix]{revtex4}
\usepackage[T1]{fontenc}
\usepackage[latin1]{inputenc}
\usepackage{graphicx}

\makeatletter

\usepackage{babel}
\makeatother
\begin{document}

\title{Improved extremal optimization for the Ising spin glass}

\author{A. Alan Middleton}

\affiliation{Department of Physics, Syracuse University, Syracuse, NY 13244}

\date{\today}

\begin{abstract}
A version of the extremal optimization (EO) algorithm introduced by
Boettcher and Percus is tested on 2D and 3D spin glasses with Gaussian
disorder. EO preferentially flips spins that are locally {}``unfit'';
the variant introduced here reduces the probability to flip previously
selected spins. Relative to EO, this adaptive algorithm finds exact
ground states with a speed-up of order $10^{4}$ ($10^{2}$) for $16^{2}$-
($8^{3}$-) spin samples. This speed-up increases rapidly with system
size, making this heuristic a useful tool in the study of materials
with quenched disorder.
\end{abstract}
\maketitle
Exploring the low temperature behavior of disordered materials, such
as spin glasses and other random magnets \cite{YoungBook}, is quite
challenging due to the very phenomena, glassy dynamics and multiple
metastable states, that are important in such materials. Scaling arguments
\cite{AndersonPond,BrayMoore84,FisherHuseSGB} indicate that many
properties of the glassy state, including the scaling of the energy
of excitations and correlation functions, can be found by studying
the ground state and its response to perturbations. Significant effort
has been invested in identifying models whose ground states can be
computed in time polynomial in the system size \cite{HartmannRieger}.
Where no polynomial-time algorithm is known, exact and heuristic methods
which take time exponential in system size are used. This enterprise
is intimately connected with concepts developed in computer science,
especially the distinction between P and NP-hard optimization problems
\cite{CompComplex}.

The Ising spin glass (ISG) is a prototypical example of a disordered
magnet. NP-hard problems such as the 3D ISG are, of course, particularly
challenging. Exact methods for the 3DISG with Gaussian bond weights
can solve $12^{3}$-spin samples with open boundary conditions \cite{PLJY2003}.
Such sizes have not proven to be sufficiently large to decide between
alternate pictures for the low-temperature behavior. Heuristic genetic
methods mix configurations and can therefore generate large scale
{}``moves'': such methods are used for samples with $14^{3}$ spins
for $\pm J$ couplings \cite{HartmannGSPMJ}. Heuristics with local
moves generally have difficulty finding the exact ground state, due
to the large barriers separating metastable states. Techniques such
as flat histogram methods \cite{WangFlat} can partially lower free
energy barriers between metastable states.

In this Communication, I study a modified version of extremal optimization
(EO) \cite{BoettcherPercus}. EO is a local search algorithm that
preferentially flips spins with low {}``fitness''. The version presented
here, {}``jaded'' extremal optimization (JEO) increases the fitness
of a spin by an amount proportional to the number of times it has
been flipped. The goal of this adjustment is to reduce the repetition
in exploring paths in configuration space, so that more possibilities
can be quickly explored. Empirically, this simple change dramatically
increases the effectiveness of the EO algorithm for finding ground
states of two- and three-dimensional spin glass samples. As exact
ground states are needed for studies of excitations and scaling, the
algorithm is, for the most part, stringently tested by demanding that
it find the ground states computed by exact methods. Both EO and JEO
take time exponential in the system size to find the exact ground
state, but the rate of growth is slower for JEO. Though JEO introduces
an extra parameter, large improvements are achieved with only modest
tuning.

\section{Extremal optimization and extended algorithm}

A principle motivation for applying EO is to explore the energy landscape
near the trial configuration by unconditionally modifying {}``unfit''
variables. Preferentially (but not exclusively) changing variables
with low fitness tends to raise the expected fitness while maintaining
large fluctuations. The algorithm differs some from traditional Monte
Carlo algorithms that conditionally select variables according to
the expected improvement. In EO, the potential moves are selected
according to their rank by fitness, rather than a Boltzmann distribution
by weight. 

A correspondence can be defined between fitness and the Hamiltonian
for the Ising spin glass \cite{BoettcherPercus}. The Hamiltonian
for spins $s_{i}$, indexed by position $i$, in a $d$-dimensional
ISG of linear size $L$ is\begin{equation}
H=-\sum_{\langle ij\rangle}J_{ij}s_{i}s_{j},\label{eq:1}\end{equation}
where $J_{ij}$ are random bond strengths each chosen with probability
$P(J_{ij})=e^{-J_{ij}^{2}/2}/\sqrt{2\pi}$ for nearest neighbor spins
with $1\le i,j\le N=L^{d}$. When $d=2$, algorithms with running
times polynomial in $N$ are available \cite{Barahona} to find the
ground state. When $d\ge3$, finding the ground state energy is NP-hard,
so that finding ground states for the worst-case choice of $J_{ij}$
is expected to take time exponential in $N$. In the context of EO,
one choice for the fitness variable $\lambda_{i}$ for a spin variable
$s_{i}$ is \begin{equation}
\lambda_{i}=\lambda_{i}^{0}\equiv s_{i}(\sum_{j\in U_{i}}J_{ij}s_{j}),\label{eq:zero}\end{equation}
where $U_{i}$ are the set of unsatisfied bonds ($s_{i}J_{ij}s_{j}<0$)
containing $s_{i}$. (Allowing for site-dependent constant shifts
$\lambda_{i}^{0}\rightarrow\lambda_{i}^{0}+\kappa_{i}$ as in Ref.~\cite{BoettcherHeap}
did not affect the comparisons here.) The configuration energy is
related to the fitness by $H=-\frac{1}{2}\sum_{i}\lambda_{i}^{0}+\sum_{ij}|J_{ij}|$.
Any increase in the fitness decreases the total energy.

Given the fitness variables $\lambda_{i}^{0}$, there are a variety
of strategies one could employ to attempt to improve the total fitness.
The simplest version of EO takes {}``greedy'' steps: the algorithm
repeatedly flips the least fit variable until a static state is achieved.
The greedy method converges quite rapidly, but in a spin glass the
convergence is to a local minimum that is generally quite far from
the optimal solution, both in configuration of the $\{ s_{i}\}$ and
often in energy per degree of freedom $H/N$. Similar greedy approaches
for decision problems such as SAT, which seeks truth assignments for
Boolean formula so that all clauses contain a true value, can be quite
successful for given ensembles of problems \cite{GreedyGood}.

An improved method, $\tau$-EO \cite{BoettcherPercus}, sorts the
spins by $\lambda_{i}$ and chooses the $m$th spin in the list with
probability proportional to $m^{-\tau}$. This favors the choice of
spins with low fitness, but allows for the occasional choice of sites
with very high fitness. Fluctuations arising from the stochastic choice
among spins with low fitness and the ranking of spins by the total
weight of broken bonds, rather than energy improvement, allow the
search to escape metastable states. It is argued \cite{BoettcherPercus}
that for large systems, the optimal choice of $\tau$ approaches $\tau=1$.

The extension considered in this paper (JEO) adjusts the fitness by
an amount proportional to the number of times $k_{i}$ that a site
$i$ has been previously chosen, that is,\begin{equation}
\lambda_{i}=\lambda_{i}^{\Gamma}\equiv\lambda_{i}^{0}+\Gamma k_{i},\label{eq:gamma}\end{equation}
where $\Gamma$ is a site-independent {}``aging'' parameter. The
variables are sorted by $\lambda_{i}^{\Gamma}$ and then selected
by rank as in $\tau$-EO. The $\tau$-EO algorithm corresponds to
the choice $\Gamma=0$. Setting $\Gamma\ne0$ reduces the probability
of selecting moves that have been flipped many times before. For configurations
near (or in) the ground state, it is favorable for \emph{some} spins
to have low fitness, in order that a number of other spins can maximize
their fitness. When $\Gamma=0$, these spins, which are actually in
their ground state orientation relative to the other spins, will be
flipped in futility. Shifting the $\lambda_{i}$ during the algorithm
also breaks the finite set of offsets between fitnesses of distinct
spins that exist at $\Gamma=0$ (due to the finite number of bond
configurations at each site). This adaptive scheme has similarities
to a variety of methods for solving problems such as SAT (satisfiability
of sets of logical constraints) that disfavor repeated selection of
the same move, such as Novelty \cite{Novelty} and variants of WALKSAT
and GSAT \cite{KautzSelman93,HSAT}. In contrast with these other
schemes, the selection process in JEO is combined with the power law
distribution for selecting ranked moves. Spin glasses with continuous
disorder differ from SAT problems as they have less local degeneracy
but also possess a global up-down symmetry, so that distinct methods
may be appropriate.

In order to select spins quickly, I used the approximate selection
method described in Ref. \cite{BoettcherHeap}. The spins are stored
in a heap structure \cite{CormenEtal} according to their current
fitness. This structure is a tree that is relatively cheap to maintain
($O(\log N)$ total cost to select a spin and update the tree). Each
spin has a parent (except for the root) and at most two children.
Each child is more fit than its parent and the root of the tree contains
the least fit spin. This structure does not guarantee any other inter-level
sorting, so that a spin $i$ that is deeper in the tree than, but
not a direct descendant of, a given spin $i'$, may have a lower fitness.
The heap structure does maintain a useful approximate sorting, though.
To select a spin to flip, a level $\ell$ is selected with probability
proportional to $2^{-(\tau-1)\ell}$ and then a random spin within
level $\ell$ is chosen. The spin at this site is then inverted. The
fitness of the neighboring spins is adjusted and the heap is updated
using standard methods \cite{CormenEtal}.

EO does not take advantage of the special structure of the 2D problem:
it is not necessary or even expected that it will find the solution
in time polynomial in the system size. Polynomial-time solvable problems
have been used to study algorithms, for example, for hard mean-field
problems \cite{RTWZ2001}. For some classes of problems, heuristics
can find solutions in polynomial time \cite{GreedyGood,SchwarzMiddleton}.
In the 2DISG, large low-energy excitations may make local algorithms
especially inefficient.

\section{Performance of the algorithm}

In this section, I compare the performance of the extended EO algorithm,
JEO, against $\tau$-EO as applied Ising spin glasses with Gaussian
disorder. When feasible, comparisons with ground states found using
exact methods provide a precise and direct test for convergence.

\emph{Two-dimensional spin glass.} The 2DISG models are on a square
lattice with $L^{2}$ spins and open boundary conditions. To determine
the 2D ground state, each sample is mapped \cite{Barahona} to a general
weighted matching problem. The matching problem for a graph is to
find a set of edges with minimal total weight such that each vertex
belongs to exactly one edge. The weighted graph for a 2DISG sample
has edges dual to the lattice bonds, with weight $|J_{ij}|$ for an
edge that crosses a bond with weight $J_{ij}$, and extra edges of
weight zero that ensure that the frustration of each plaquette is
maintained: unfrustrated (frustrated) plaquettes give an even (odd)
number of the bonds dual to the edges of the plaquette in the matching.
To find the minimum weight matching and hence the ground state energy
for a 2DISG sample, I used the Blossom IV algorithm developed by Cook
and Rohe \cite{Blossom}. 

The exact ground state energy of each 2DISG sample was input to the
$\tau$-EO and JEO codes. When the heuristic codes found this energy,
the codes terminated. The primary results from these computations
were the distributions of the running times, measured in number of
spin flips, to find the true ground state. The time to solution is
a function of both the seed used to generate the sample and an independent
{}``algorithm seed'' used to generate the random initial configuration
and to select spin flips. In a given sample, the distribution of times
to find a ground state was roughly Poissonian. This suggests that
restarting the algorithm with different initial configurations or
seeds for selecting flips does not significantly decrease the mean
running time. This conclusion was consistent with empirical trials
of restarting the algorithm: the algorithm does not get stuck in history
dependent traps. Given a sample $k$, the median $t_{m}^{k}$ of the
running time was estimated from the solution time for 100 algorithm
seeds. The results reported here are for $\overline{t}_{m}$, the sample
mean of $t_{m}^{k}$. The $\Gamma=0$ data is in agreement
with  previously results for $\tau$-EO, with $\overline{t}_{m}$
minimal at $\tau\approx1.5$.

\begin{figure}
\includegraphics[%
  width=1.0\columnwidth]{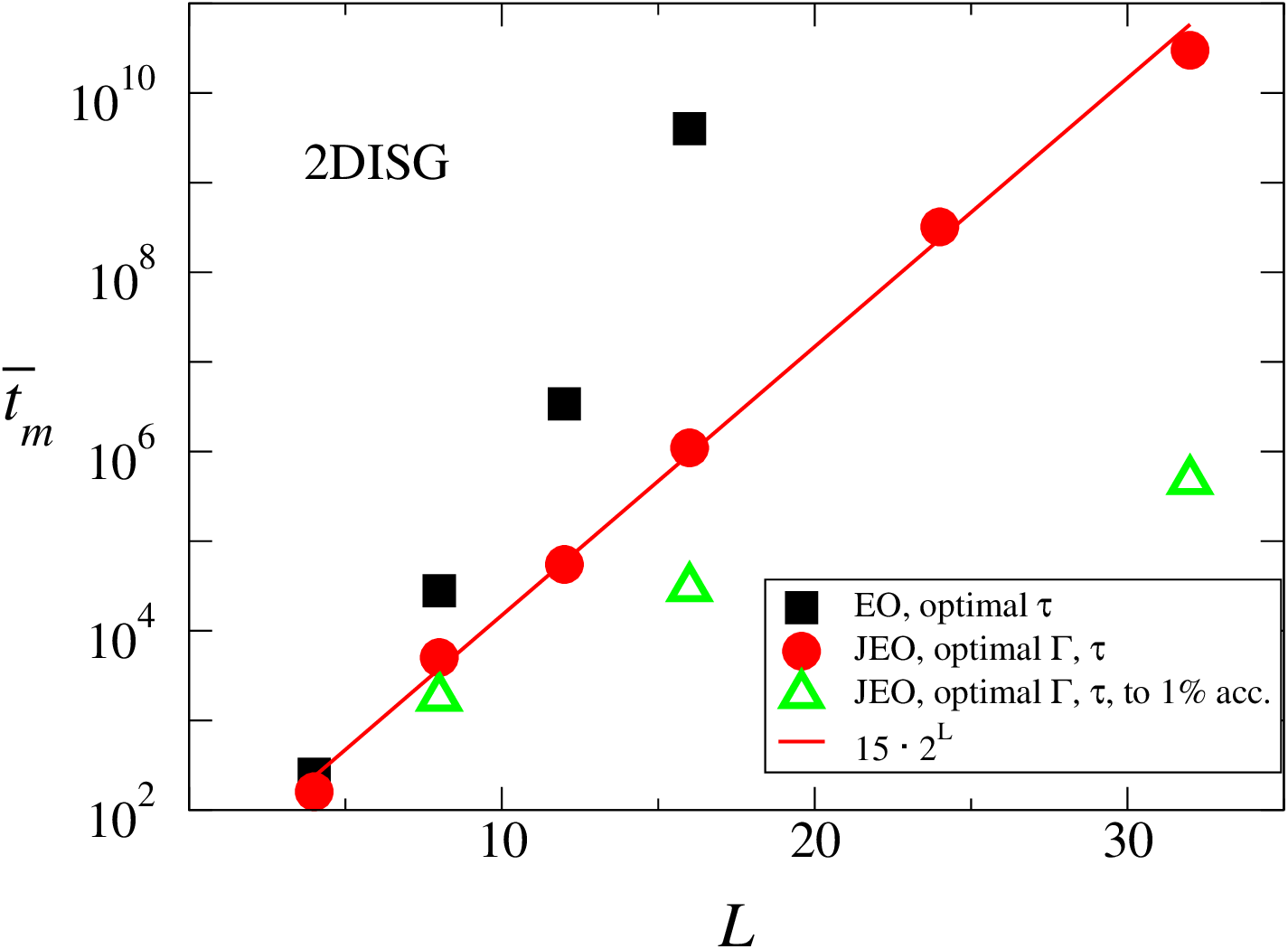}

\caption{\label{cap:Summary}Plot of $\overline{t}_{m}$, the sample mean
of the median time to find the ground state , measured in spin flips,
using $\tau$-EO (squares) and JEO (circles), for the 2DISG with optimal
$\tau$ and, for JEO, $\Gamma$. The triangles indicate the same measure
of time to find the ground state energy to within 1\% accuracy. The
line shows, for comparison, a running time exponential in $L$, $\overline{t}_{m}=15\cdot2^{L}$,
consistent with the results for JEO. The uncertainties are comparable
to the symbol size.}
\end{figure}

\begin{figure}
\includegraphics[%
  width=1.0\columnwidth]{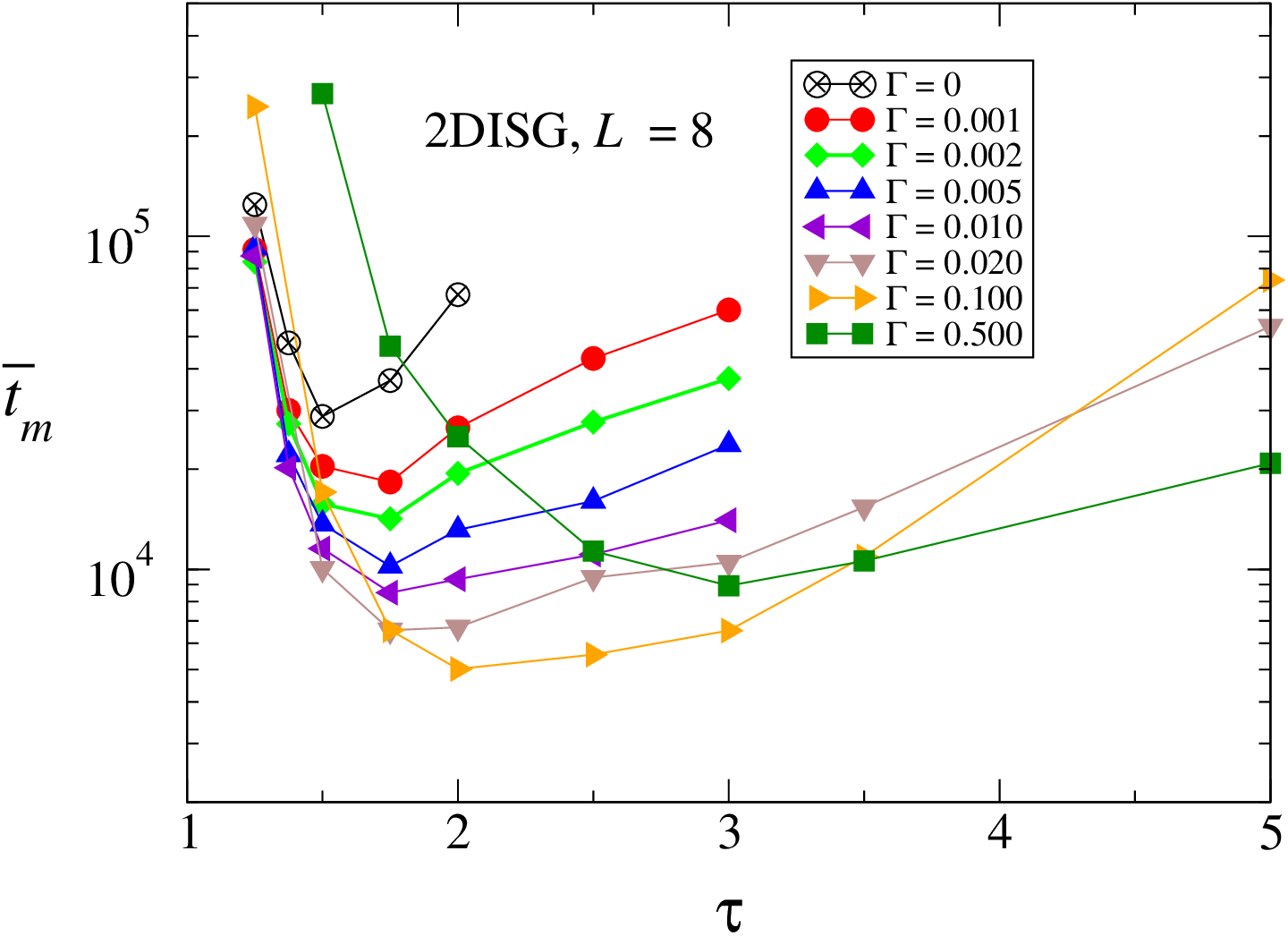}

\caption{\label{cap:vstau}Plot of $\overline{t}_{m}$ for 2DISG samples of
size $L=8$, for $\Gamma$ ranging from $\Gamma=0$ (i.e., $\tau$-EO)
through $\Gamma=0.5$, as a function of the power law for rank selection,
$\tau$. For clarity, the error bars, which are of order 10\% of the
values for all points, are not shown. The solid lines are added only
to group the points. Choosing $\Gamma\approx0.1$ and $\tau\approx2.0$
minimizes the run time.}
\end{figure}

The results for the mean solution time $\overline{t}_{m}$ for optimal
$\tau$ and $\Gamma$ are summarized in Fig.~\ref{cap:Summary}.
As suggested by the data plotted in Fig.~\ref{cap:vstau}, $\overline{t}_{m}$
is not very sensitive to the exact choice of parameters, as long as
$\tau$ is in the range $1.5<\tau<2.5$ and the optimal $\Gamma$
(on the order of $10^{-3}$ to $10^{-1}$) is found to within a factor
of about 2, for the sizes studied here. The best running times for
$\tau$-EO grow much more rapidly than those for JEO. For $L=16$,
JEO is of the order $10^{4}$ times faster than $\tau$-EO. Extrapolation
suggests that the advantage of JEO increases significantly with $L$.
For comparison, an exponential dependence $\overline{t}_{m}=15\cdot2^{L}$
is shown in Fig.~\ref{cap:Summary}. This function does a good job
of describing the JEO data for $L=4$ through $L=32$. In separate
runs, for comparison, the heuristic algorithm was terminated when
the energy was within 1\% of the exact ground sate energy. These approximate
solutions were found much more rapidly than exact solutions ($\approx10^{5}$
times faster for $L=32$).

\begin{figure}
\includegraphics[%
  width=1.0\columnwidth]{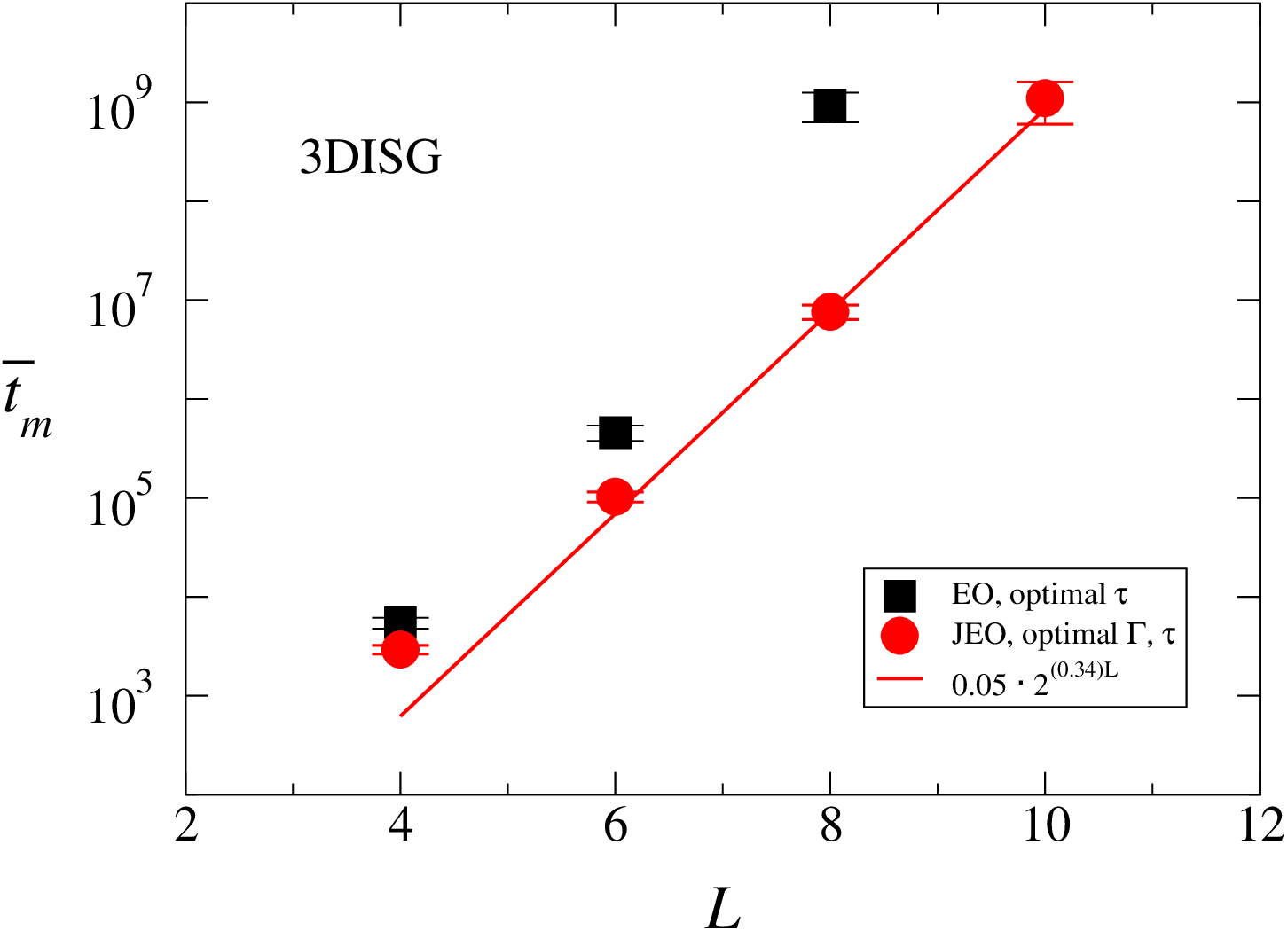}

\caption{\label{cap:3D}Plot of the sample average of the median running times
for $\tau$-EO (squares) and JEO (circles) for the Gaussian Ising
spin glass on a cubic lattice. The algorithm terminated when 8 of
the minimal record energies agreed among 10 parallel samples. The
parameter $\tau$ was fixed for JEO at a near-optimal $\tau=1.7$
and near-optimal values of $\Gamma=0.1,0.1,0.05$ for $L=4,6,8$,
respectively, were used. The gain for JEO over $\tau$-EO is approximately
a factor of 100 at $L=8$. The line shows $\overline{t}_{m}=0.05\cdot2^{3.4\cdot L}$,
for a rough comparison.}
\end{figure}

\emph{Three-dimensional spin glass.} A similar comparison was carried
out for 3DISG samples with Gaussian disorder. The $L^{3}$ spins in
the 3DISG samples lie on a cubic lattice with periodic boundary conditions.
For 3DISG samples of size up to $6^{3}$, the spin glass server at
the University of Köln \cite{SGserver} (which applies branch-and-cut
\cite{HartmannRieger}) was used to generate exact solutions. The
termination condition of the algorithm was modified, as exact ground
states for the larger samples were not readily available. All samples
were simulated in parallel with $n=10$ algorithm seeds. When the
minimal record energy for eight (8) of the samples were identical,
the algorithm was terminated. This criterion produced configurations
equal to the exact solutions for all $L=4,6$ samples (45 at each
size). This suggests that true ground states were found with a high
probability for $L=8$ and possibly also $L=10$. The summary results
are plotted in Fig.~\ref{cap:3D}. Given the termination criterion,
JEO was of the order of $10^{2}$ times faster than $\tau$-EO in
converging to a potential solution for $L=8$ samples. Very roughly,
$L=6$ samples were solved in $\approx10\,\textrm{s}$ on average
both on the Köln spin glass server (a 400 MHz Sun Ultra) and using
JEO (on a 1 GHz Intel P5). Further studies would be needed to provide
better estimates of the confidence in the ground states and how to
improve such confidence.

\section{Discussion}

JEO extends the extremal optimization algorithm of Boettcher and Percus
by adaptively reducing the frequency of flipping previously selected
spins. As a local move can lead to avalanche-like behavior, due to
induced changes in the fitness of neighbors, this modification also
reduces the frequency of flipping larger domains. This extension of
EO does add a parameter, the aging parameter $\Gamma$. However, a
near-optimal value for $\Gamma$ for each problem type at a given
size can be found quickly and less tuning of the parameter $\tau$
is required than for $\tau$-EO.

One possible avenue of exploration is to check whether avalanche regions
correspond to important domains or excitations in the sample. Possible
modifications of JEO include using a selection distribution with sharp
cutoffs \cite{FranzHoffmann}, rather than power-law distributions.
Other schemes for reducing the fitness of frequently repeated moves
could be considered, such as modifying the fitness using non-linear
functions of the number of flips at a spin.

Regardless of the exact details of the role of domains and possible
improvements, empirical testing shows that the aging of the spins
during state-space exploration greatly reduces the time for EO to
find the ground state of the ISG in two and three dimensions. Though
the 2D model was used to make a precise comparison with exact results,
the exponential equilibration times for the 2DISG using extremal optimization
are consistent with those that would be seen for an NP-hard optimization
problem with a similar local solution strategy. It may be useful to
use an algorithm like JEO to locally improve the configurations formed
by whole sample crossover in genetic algorithms \cite{MartinHoudayerRenorm}.
As exact solutions for small samples can be found with confidence
in a relatively small number of steps, in machine time very similar
to that for branch-and-cut, this simple algorithm also provides a
very convenient way to study small 3D samples. 

I thank Stefan Boettcher for discussions and comments. The Köln spin
glass server was quite useful for this work. I thank the Kavli Institute
for Theoretical Physics and the Schloss Dagstuhl Seminar (03381) for
their hospitality. This work was supported in part by the National
Science Foundation (grants DMR-0109164 and DMR-0219292).

\end{document}